\documentclass[12pt]{iopart}

\usepackage{amssymb}
\usepackage{amsfonts}
\usepackage{bm}
\usepackage{epsfig}
\usepackage{epsf}                          
\usepackage[]{graphicx}  

\usepackage{color}

\renewcommand {\Re} {{\rm Re}}
\renewcommand {\Im} {{\rm Im}}

\renewcommand {\d} {{\rm d}}
\renewcommand {\i} {{\rm i}}

\newcommand {\ee} {{\rm e}}

\newcommand {\bfe} {{\bf e}}
\newcommand {\bfd} {{\bf d}}
\newcommand {\bfn} {{\bf n}}
\newcommand {\bfp} {{\bf p}}
\newcommand {\bfr} {{\bf r}}

\newcommand {\bfE} {{\bf E}}
\newcommand {\bfD} {{\bf D}}

\newcommand {\E} {\varepsilon}
\newcommand {\om} {\omega}

\newcommand {\tom} {\tilde{\omega}}

\newcommand {\calF} {{\cal F}}
\newcommand {\calG} {{\cal G}}
\newcommand {\calM} {{\cal M}}
\newcommand {\calU} {{\cal U}}

\newcommand {\Ia} {I_{\rm a}}

\begin{document}
\jl{2}


\title[Photoionization of a strongly polarizable target]
{Photoionization of a strongly polarizable target}

\author{A V Korol$^{1,2}$ and A V Solov'yov$^1$}

\address{$^1$ Frankfurt Institute for Advanced Studies, 
Johann Wolfgang Goethe-Universit\"at, 
Ruth-Moufang-Str. 1, 60438 Frankfurt am Main, Germany}
\address{$^2$ Department of Physics, 
St.Petersburg State Maritime Technical University, 
Leninskii prospect 101, St.~Petersburg 198262, Russia}

\eads{korol@fias.uni-frankfurt.de, solovyov@fias.uni-frankfurt.de}

\date{today}

\begin{abstract}
{
We demonstrate that the angular distribution of photoelectrons 
from a strongly polarizable target exposed to a laser field   
can deviate noticeably from the prediction of  conventional theory. 
Even within the dipole-photon approximation
the profile of distribution is modified due to the action of the field 
of alternating dipole moment induced at the residue by the laser field. 
This effect, being quite sensitive to the dynamic polarizability 
of the residue and to its geometry,
depends also on the intensity and frequency of the laser field.  
Numerical results, presented for sodium cluster anions,
demonstrate that dramatic changes to the profile occur
for the photon energies in vicinities of the plasmon resonances,
where the effect is enhanced due to  the increase in the 
residue polarizability.
Strong modifications of the characteristics of a single-photon 
ionization process can be achieved by applying laser fields of comparatively 
low intensities $I_0 \sim10^{10}-10^{11}$  W/cm$^2$.
}
\end{abstract}

\section{Introduction \label{Introduction}} 

In this paper we demonstrate that the profile of angular distribution of 
electrons emitted in the process of a single-photon ionization 
from a strongly polarizable target 
(in particular, a metal cluster anion)
exposed to a laser field can be noticeably modified 
due to the action of the field $\calU(\bfr,t)$ of alternating dipole moment 
$\bfD(t)$ induced at the residue by the laser field.
The field $\calU(\bfr,t)$ acts on the escaping electron and, 
being dependent on the polarization vector of the laser field,
brings additional dependence of the cross section on the emission angle.
The degree of this dependence is determined by the magnitude of $\bfD(t)$ 
which, in turn, is proportional to the residue's dynamic polarizability.
Therefore,  for a target with large polarizability the effect 
can be very pronounced even for laser fields of low intensities.

The effect of the laser-field--induced dipole moment in atomic photoionization
process was discussed in \cite{GolovinskiKiyan1990} (where one finds also 
the references to the earlier works by the authors) and, more recently, in 
\cite{ChernovKiyanHelmZon_2005}. 
Both papers were devoted to the study of {\it multiphoton} 
detachment of electrons from atomic negative ions. 
In \cite{GolovinskiKiyan1990} the process was analyzed by means of 
the perturbation theory. 
Numerical analysis was carried out for halogen and alkali metal anions,   
and it demonstrated, that the induced dipole moment influences the magnitude
of total cross section of the many-photon detachment.
The paper \cite{ChernovKiyanHelmZon_2005} dealt with the process of a
{\it strong-field} many-photon detachment of an atomic anion.
The non-perturbative adiabatic theory, developed in 
\cite{GribakinKuchiev_1997}, was modified to account for the 
field of the induced dipole moment.
The numerical calculations, performed for Rb$^{-}$ and Cs$^{-}$,
showed the increase of the detachment rate and essential modification of 
the angular distribution.
 
In our paper we investigate the influence of the induced dipole moment
on the angular and  spectral distributions of photoelectrons in the 
process of {\it single-photon} ionization (or detachment) 
occurring in the {\it weak-field} regime.
Numerical results are presented for sodium cluster anions Na$_N^{-}$ 
with $N=10^1\dots10^2$.
The corresponding residue clusters are characterized by large values of 
the dynamic dipole polarizability $\alpha(\om)$, which in the vicinity 
of plasmon resonances,
i.e. at $\om = 2\dots3$ eV, can be as large as $10^4\dots 10^5$ a.u.,
which by far exceed the maximum values of $\alpha(\om)$
of individual atoms (e.g., $\alpha(\om) \approx 400$ a.u.
for Cs at the photon energy 0.47 eV \cite{ChernovKiyanHelmZon_2005}).
  
We demonstrate that dramatic changes to the profile of angular 
distribution occur in vicinities of the plasmon resonances
and for comparatively low intensities $\sim10^{10}-10^{11}$  W/cm$^2$
of the laser field.
These are exactly the ranges used in recent experimental studies 
\cite{KostkoEtAl_2007,Kostko_disser} of the photodetachment from
various metal cluster anions carried out by means of  
angle resolved photoelectron spectroscopy.
Therefore, the experimental test of the predicted effects seems to be 
feasible.

The atomic system of units, $e=m=\hbar=1$ is used throughout the paper.

\section{Theoretical framework
\label{Framework}} 

\subsection{Setting the  problem
\label{Problem}} 

Consider the process of photoionization 
of a strongly polarizable target 
(a cluster, a fullerene, a complex molecule, etc) 
by spatially homogenous linearly polarized laser field
\begin{eqnarray}
\bfE(t) 
=
\bfE_0\cos\om t 
=
{1\over 2}
\left(\bfE_0\ee^{-\i\om t}
+
\bfE_0\ee^{\i\om t}\right)
\,.
\label{Problem.1}
\end{eqnarray}
We assume, that the field wavelength $\lambda$ greatly exceeds
the characteristic size $R$ of the target, $\lambda \gg R$.
Thus, the process can be treated within the 
dipole-photon limit.

The laser field ionizes the target and polarizes the residue.
Therefore, in addition to other fields 
(the static potential of the residue, the long-range polarizational 
potential)
the escaping photoelectron feels the field due to the oscillating 
dipole moment $\bfD(t)$ induced at the residue: 
\begin{eqnarray}
\calU(\bfr,t) = -{ \bfD(t){\bfr} \over r^3}\,.
\label{Problem.2}
\end{eqnarray}

To simplify the consideration we assume, that the residue 
is spherically symmetric, so that its tensor of polarizability 
contains only a scalar part which is notated as $\alpha(\om)$.
Then, the moment $\bfD(t)$ reads
\begin{eqnarray}
\bfD(t)
 = 
\bfd\,\ee^{-\i\om t}
+
\bfd^{*}\,\ee^{\i\om t},
\label{Problem2.1}
\end{eqnarray}
where $\bfd \equiv \bfd(\om) = \alpha(\om)\, \bfE_0/2$.
The field of the induced dipole can be written as
follows:
\begin{eqnarray}
\calU(\bfr,t) 
=
U(\bfr)\,\ee^{-\i\om t}
+
U^{*}(\bfr)\,\ee^{\i\om t},
\label{Problem2.3a}
\end{eqnarray}
with $U(\bfr) = \bfd\bfr/ r^3$.
This field is dependent on the direction of $\bfE_0$, therefore,
the angular distribution of photoelectrons acquires additional 
dependence on the polarization of the laser field.
The degree of this dependence is determined by the magnitude of $\bfD(t)$,
which is proportional to the residue dynamic
dipole polarizability.
Hence, for a target with large polarizability the modification
of the angular distribution due to 
the induced dipole moment can become noticeable even for 
a laser field of a comparatively low intensity.

\subsection{The perturbation series
\label{Perturbation}} 

Let us clearly state the conditions which we impose on the parameters
of the laser field.
These include the following inequalities:
\begin{eqnarray}
{\rm (I)}\ E_0 \ll E_{\rm int},
\qquad
{\rm (II)}\ \E \gg {E_0^2 \over 4\om^2},
\qquad
{\rm (III)}\ 
\om > |\E_{\rm i}|\,.
\label{conditions}
\end{eqnarray}
The first condition implies that the laser field strength is much 
smaller than the strength $E_{\rm int}$ of the internal field in the target.
This condition allows one to ignore the laser field when 
constructing the wavefunction of the bound electron with the binding energy
$\E_{\rm i} < 0$.
Numerical results presented in section \ref{NumericalResults} 
refer to the laser field strength 
$E_0 \sim 10^{-3}\, {\rm a.u.} \sim 10^6$ V/cm.  

The second condition states that the energy $\E$ of the escaping 
electron is large compared to the ponderomotive shift due to 
the wiggling in the laser field. 
Owing to this condition we do not ``dress'' the photoelectron 
wavefunction with the laser field. 

The last inequality in (\ref{conditions}) indicates that 
the photon energy exceeds the ionization threshold of the bound 
electron.
Therefore, the process of a 
{\it single-photon} ionization is allowed 
for which the energy conservation law reads
\begin{eqnarray}
\E = \E_{\rm i} + \om\,.
\label{M0.3}
\end{eqnarray}

Combining the conditions (I) and (III) one demonstrates that the following 
inequality is met: $\gamma \equiv \om \kappa/ E_0 \gg 1$ where
$\gamma$ is a so-called Keldysh parameter, and 
$\kappa = (2|\E_{\rm i}|)^{1/2}$.
The limit $\gamma\gg 1$ defines a weak field regime 
(see, e.g. \cite{GribakinKuchiev_1997}).
Thus, in this paper we investigate the influence of the induced dipole moment
$\bfD(t)$ on the angular and  spectral distributions of 
photoelectrons in the process of
{\it single-photon} ionization (or detachment) 
occurring in the weak field regime.
Earlier the role of $\bfD(t)$ was studied in the process of many-photon 
detachment  (i.e., when $\om \ll |\E_{\rm i}|$) 
from atomic negative ions. 
It was done in the weak-field  \cite{GolovinskiKiyan1990} and
in the strong-field \cite{ChernovKiyanHelmZon_2005} limits.   

Let us note, that the weak field condition (II) does not immediately 
imply the applicability of the perturbative approach 
with respect to $\calU(\bfr,t)$.
Indeed, from (\ref{Problem.2})-(\ref{Problem2.3a}) follows that
$|\calU(\bfr,t)|\sim E_0 |\alpha(\om)|/r^2$
exceeds $E_0$ by a factor $|\alpha(\om)|/r^2$.
For a strongly polarizable target this factor
can be large enough in a wide range of radial distances $r>R$, 
and this might lead to a non-perturbative treatment of the action of 
$\calU(\bfr,t)$.

To carry out a quantitative analysis we first postulate that under
certain conditions the field (\ref{Problem2.3a}) can be considered
as a time-dependent perturbation which modifies the wavefunction of 
the escaping electron.
The criterion of applicability of this approach will be
formulated in the course of calculations. 

\begin{figure}[h]
\centering
\includegraphics[clip,scale=0.5]{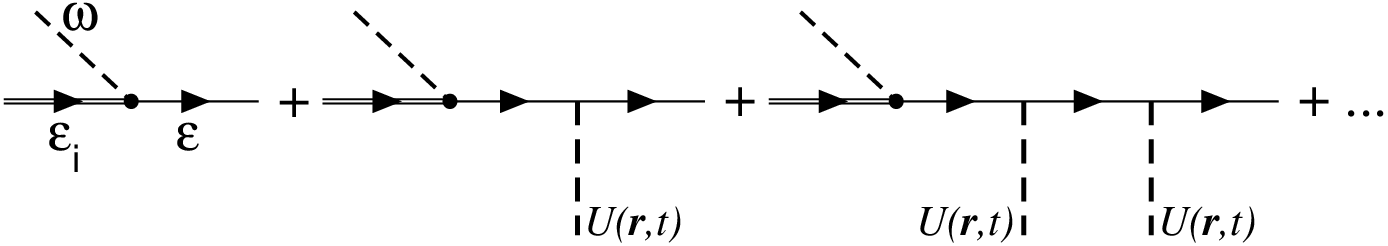}
\caption{
Diagrammatical representation of a single-photon 
transition amplitude 
in a form of perturbation series in $\calU(\bfr,t)$ 
(indicated with the dashed vertical lines).
The double line represents the initial bound state.
The solid line stands for the escaping electron whose
wavefunction is ``dressed'' with all static fields.
}
\label{diagram.fig}
\end{figure}

The matrix element $\calM$, which  describes a dipole-photon 
transition of the electron from the initial bound state
$\psi_i(\bfr,t) = \psi_i(\bfr)\ee^{-\i\E_{\rm i} t}$ 
to the final state $\psi_f(\bfr,t) = \psi^{(-)}_{\bfp}(\bfr)\ee^{-\i\E t}$  
with the asymptotic momentum $\bfp$ and energy $\E=p^2/2$, 
can be written in the following form:
\begin{eqnarray}
\calM = \calM_0 + \calM_1 + \calM_2 + \dots
\label{Problem.3}
\end{eqnarray}
The right-hand side of this equation represents the power series in 
$\calU(\bfr,t)$.
The structure of the terms of the series is illustrated
by diagrams in figure \ref{diagram.fig}.

The zeroth order term $\calM_0$ 
describes the direct process of photoionization standing
for the  matrix element of the 
operator $\bfE_0\bfr \exp(-\i\om t)/2$: 
\begin{eqnarray}
\calM_0 = {1\over 2}
\int_{-\infty}^{\infty}
\d t \,
\psi_f^{*}(\bfr,t) \,
\bfE_0\bfr \, \ee^{-\i\om t}\,
\psi_i(\bfr,t) 
=
2\pi\delta(\E - \E_{\rm i} - \om)\, M_0,
\label{M0.1}
\end{eqnarray}
where
\begin{eqnarray}
M_0 
=
{1\over 2}\int \d\bfr \, \psi^{(-)*}_{\bfp}(\bfr)\,(\bfE_0\bfr)\, 
\psi_i(\bfr)\,.
\label{M0.2}
\end{eqnarray}
We assume that the wavefunction  $\psi^{(-)}_{\bfp}(\bfr)$ 
(whose asymptotic form is 'the plane wave + the incoming wave') 
corresponds to the state ``dressed'' with all {\it static} fields.
The delta-function on the right-hand side of (\ref{M0.1}) 
expresses the energy conservation law (\ref{M0.3}).

The terms $\calM_n$ with $n\geq 1$ are the corrections to the transition  
amplitude due to the $n$-times interaction of the photoelectron
with $\calU(\bfr,t)$.
Since this field explicitly depends on $t$ (see (\ref{Problem2.3a})) 
then in each act of the interaction the electron energy is 
changed (increased or decreased) by $\om$.
As a result, not all terms on the right-hand side of (\ref{Problem.3})
represent the correction to the direct amplitude $\calM_0$, which 
implies the validity of (\ref{M0.3}).
For example, evaluating $\calM_1$ (the second 
diagram in figure \ref{diagram.fig}) one finds, that it contains 
the terms proportional to 
$\delta(\E - \E_{\rm i})$ and to $\delta(\E - \E_{\rm i} - 2\om)$, which 
conflict with the conservation law (\ref{M0.3}).
Therefore, the term $\calM_1$ must be ignored when constructing 
the correction to  $\calM_0$.

The lowest-order correction originates from the amplitude $\calM_2$
(the last diagram in figure \ref{diagram.fig}),
which contains the terms proportional to $\delta(\E - \E_{\rm i} - \om)$.
Applying the standard technique of the time-dependent perturbation theory,
one derives the following expression for this contribution:
\begin{eqnarray}
\left.\calM_2\right|_{\E = \E_{\rm i} + \om} 
=
2\pi\delta(\E-\E_{\rm i}-\om)M_2
\label{M2.2}
\end{eqnarray}
where
\begin{eqnarray}
\fl
M_2
&=
{|\alpha(\om)|^2\,E_0^2 \over 8}
\int\!\!\int 
\d\bfr_1\, \d\bfr_2\,
\psi_{\bfp}^{(-)*}(\bfr_2)
{\bfe_0\bfr_2 \over r_2^3} \,
\Biggl(
G_{\E_{\rm i}}(\bfr_2,\bfr_1) 
+ 
 G_{\E+\om}(\bfr_2,\bfr_1) 
\Biggr)\,
{\bfe_0\bfr_1 \over r_1^3}
\nonumber\\
\fl
&\times
\int \d\bfr\,
G_{\E}(\bfr_1,\bfr)\, 
(\bfE_0\bfr) \,
\psi_i(\bfr).
\label{Ampl.3}
\end{eqnarray}
Here $G_E(\bfr^{\prime},\bfr)$ is 
the Green's function of the photoelectron and
$\bfe_0=\bfE_0/E_0$ is the unit vector of the laser field polarization.

Hence, within the second second order perturbation theory in
 $\calU(\bfr,t)$, 
the photoionization amplitude is written as:
\begin{eqnarray}
\left.\calM_2\right|_{\E = \E_{\rm i} + \om} 
\approx
2\pi\delta(\E-\E_{\rm i}-\om)
\Bigl( M_0 + M_2 \Bigr)
\label{Ampl.1}
\end{eqnarray}

Using the amplitude (\ref{Ampl.1}) one can write the following
expression for the spectra-angular distribution:
\begin{eqnarray}
{\d \sigma \over \d \Omega} 
=
{p \over 4\pi^2c}\,
\Bigl|M_0+M_2\Bigr|^2 
\approx
{\d \sigma_0 \over \d \Omega} 
\Bigl(1 + K(\om,\theta)\Bigr).
\label{CS.3}
\end{eqnarray}
Here  
\begin{eqnarray}
{\d \sigma_0 \over \d \Omega} 
=
{p \over 4\pi^2c}\,
\Bigl|M_0\Bigr|^2
\label{CS.4}
\end{eqnarray}
stands for the differential cross section calculated without the 
influence of the induced dipole moment.
Within the dipole-photon approximation the differential 
cross section $\d \sigma_0 / \d \Omega$ can be presented in the 
well-known general form  (see, e.g., \cite{Drake2006}):
\begin{eqnarray}
{\d \sigma_0 \over \d \Omega} 
=
{\sigma_0 \over 4\pi}\,
\Bigl(1 + \beta(\om)P_2(\cos\theta)\Bigr),
\label{CS.4a}
\end{eqnarray}
where $\sigma_0\equiv \sigma_0(\om)$ is the total cross 
section of photoionization, $\theta$ stands for the angle 
between $\bfp$ and $\bfe_0$, $P_2(\cos\theta)$ is the Legendre 
polynomial of the second order. 
The quantity $\beta(\om)$ stands for the angular asymmetry parameter 
\cite{CooperZare1968}. 
The factor $1 + \beta(\om)P_2(\cos\theta)$ defines the profile
of the angular distribution.

Equation  (\ref{CS.3}) indicates, that the profile is modified by the factor
$1+K(\om,\theta)$, when the field (\ref{Problem2.3a}) is taken into account.
The correction term $K(\om,\theta)$ is defined as follows:
\begin{eqnarray}
K(\om,\theta)
=
2\,{\rm Re}\, {  M_2 \over M_0}.
\label{CS.5}
\end{eqnarray}

Formula (\ref{CS.3}) was obtained by means of the perturbation theory.
Thus, it is implied that the absolute value of the correction term
satisfies the condition $|K(\om,\theta)| < 1$. 
If otherwise, the field  $\calU(\bfr,t)$ must be accounted for
in all orders of the perturbation series (\ref{Problem.3}).

It follows from (\ref{Ampl.3}), that $K(\om,\theta)$ is proportional to 
the squared induced dipole moment,
$K\propto |\bfd|^2 =|\alpha(\om)|^2\,E_0^2$. 
Therefore, for a target of a small polarizability,
one has to apply very intensive laser fields
in order to achieve a noticeable correction to the photoelectron 
angular distribution, i.e. $|K(\om,\theta)| \sim 1$. 
On the contrary, for a strongly polarizable target 
(e.g., a metallic cluster or/and its anion)
a strong effect can be expected for laser fields of a moderate strength.
In section \ref{NumericalResults} we demonstrate, that dramatic change
of the angular distribution profile can occur for 
$E_0 \sim 10^6$ V/cm.

It is important to note, that the correction term (\ref{CS.5})
depends on the ratio of the amplitudes $M_2$ and $M_0$.
Thus, one can expect that it is less sensitive to the approximation 
chosen to describe the photoelectron
wavefunction than the cross section $\d \sigma_0/ \d \Omega$.
Therefore, to estimate the influence of the target polarization one can
calculate $K(\om,\theta)$ within the framework of a comparatively simple 
 approximation, which is described below.

\subsection{Analytic expression for $K(\om,\theta)$ 
\label{AnalyticExpression}} 

In this section we derive an approximate expression for the 
correction term $K(\om,\theta)$ using a simple, although realistic and
physically clear, model.

To start with let us assume, that the ionized target is 
a spherically-symmetric anion.
Then, firstly, the ground state wavefunction $\psi_i(\bfr)$ 
does not depend on the angles of the position vector:
$\psi_i(\bfr)=\psi_i(r)$.
Secondly, the wavefunction $\psi_{\bfp}^{(-)}(\bfr)$ of the photoelectron
does not contain the contribution of a long-range Coulomb field of 
the residue.
These assumptions, being not of a principal nature, allow one to
simplify the intermediate algebra.  

Making use of the partial wave expansion 
\begin{eqnarray}
\psi_{\bfp}^{(-)}(\bfr)
= 
{4\pi \over pr}
\sum_{lm}
\i^{l}\,\ee^{-\i\delta_l(p)} \,Y_{lm}^{*}(\bfn_\bfr)Y_{lm}(\bfn_\bfp)\,
\chi_{pl}(r),
\label{M0.1a}
\end{eqnarray}
where $Y_{lm}(\bfn)$ are the spherical harmonics,
$\delta_l(p)$ are the scattering phaseshift,
and  $\chi_{pl}(r)$ are the radial wavefunctions,
one writes the amplitude $M_0$ in the following form:
\begin{eqnarray}
M_0 
=
\left.
-2\pi\i\, {\bfE_0 \bfp\over p^2}\,\ee^{\i\delta_l(p)} \,
\int_0^{\infty} \d r\,r^2\, 
\chi_{pl}(r)\, \psi_i(r)
\right|_{l=1}\,.
\label{M0_Born.1}
\end{eqnarray}

Another simplification is based on the approximation that
the field $\calU(\bfr,t)$ acts on the escaping electron only outside 
the target, i.e. at the distances $r > R$.
Additionally, one can assume, that the wavefunction $\psi_i(r)$ is 
concentrated in the interior of the target, i.e. at $r<R$.
Therefore, when evaluating the right-hand side of (\ref{Ampl.3}) one 
takes into account that $r_1, r_2 > R >r$.
This allows one to relate $M_2$ to the amplitude $M_0$.
To do this let us approximate the Green's function $G_{\E}(\bfr_1,\bfr)$
as follows:
\begin{eqnarray}
\fl
G_{\E}(\bfr_1,\bfr) 
=
2 \i p
\sum_{lm} 
\calG_{\E l}(r_1,r)\,
Y_{lm}(\bfn_1) Y_{lm}^{*}(\bfn)
\approx
{2 \i  \over p} 
\sum_{lm} 
{X_{pl}(r_{1}) \chi_{pl}(r) \over r_{1}r}\,
Y_{lm}(\bfn_1) Y_{lm}^{*}(\bfn).
\label{LongDistance.2}
\end{eqnarray}
Here $r_{>}/ r_{<}$ is the largest/smallest of $r_1, r$,
$X_{pl}(r)$ stands for the irregular solutions of the radial 
Schr\"{o}dinger equation for the energy $\E=p^2/2$.
The quantity 
$g_{\E l}(r_1,r) =X_{pl}(r_{>})\chi_{pl}(r_{<})/p^2r_{>}r_{<}$ 
is the exact radial Green's function corresponding to the 
multipolarity $l$.
The first equality in (\ref{LongDistance.2}) states the partial-wave
expansion of the Green's function.
The approximate equality  is based on 
the assumption $r_1>r$. 

Using (\ref{LongDistance.2}) in (\ref{Ampl.3}) one notices
that the radial integral over $r$ reduces to that
from (\ref{M0_Born.1}).
Therefore, the amplitude $M_2$ acquires the form:
\begin{eqnarray}
M_2
\approx
- \gamma(\om, \theta) M_0,
\label{LongDistance.7}
\end{eqnarray}
with
\begin{eqnarray}
\fl
\gamma(\om, \theta)
&=
 |\alpha(\om)|^2 E_0^2 \, 
{\ee^{-\i\delta_1(p)}\over 8\pi}\,\cos^{-1}\theta
\nonumber\\
\fl
&\times
\int\!\!\int\!
\d\bfr_1 \d\bfr_2\,
\psi_{\bfp}^{(-)*}(\bfr_2)
{\bfe_0\bfn_2 \over r_2^2} 
\Biggl(
G_{\E_{\rm i}}(\bfr_2,\bfr_1) 
+ 
 G_{\E+\om}(\bfr_2,\bfr_1) 
\Biggr)
{(\bfe_0\bfn_1)^2 \over r_1^3}
 X_{p1}(r_1).
\label{LongDistance.8}
\end{eqnarray}
Here the integration is carried out 
over the spatial region ${r_1,r_2>R}$.

The correction term (\ref{CS.5}) is related to $\gamma(\om, \theta)$
as $K(\om,\theta) = -2{\rm Re}\,\gamma(\om, \theta) $.

Let us comment on the evaluation of the right-hand side of 
(\ref{LongDistance.8}). 
Introducing the expansion (\ref{M0.1a}) and representing the Green's 
functions 
$G_{\E_{\rm i}}(\bfr_2,\bfr_1)$ and $G_{\E+\om}(\bfr_2,\bfr_1)$ 
in terms of the partial-wave series similar to (\ref{LongDistance.2}), 
one carries out the angular integration by means of the standard 
methods (see, e.g., \cite{BMX}) and arrives at:
\begin{eqnarray}
\fl
K(\om,\theta) 
&= 
-2\Re\, \gamma(\om, \theta)
\nonumber\\
\fl
&=
-{1\over 3}
|\alpha(\om)|^2 E_0^2 
\sum_{ll^{\prime}}
 \left[
\delta_{l1}
\left(
\delta_{l^{\prime}0}
+ 
{4  \over 5}\delta_{l^{\prime}2}
\right)
-
{6 \over 5}
\delta_{l3}\delta_{l^{\prime}2}
{P_3(\cos\theta) \over \cos\theta}
\right]
\Im\, B_{ll^{\prime}}.
\label{LongDistance1.5}
\end{eqnarray}
The Kronecker symbols $\delta_{ij}$, which reduce the allowed values of 
the orbital momenta to  $l=1,3$ and $l^{\prime}=0,2$, reflect the dipole
selection rules applicable to each of the vertices in the last 
diagram in figure \ref{diagram.fig}. 
The ratio ${P_3(\cos\theta) / \cos\theta}$ reduces to 
$(5P_2(\cos\theta) - 2)/3$ 
with the help of the recurrence relation for the Legendre polynomials. 

The quantities  $B_{ll^{\prime}}$ stand for the following radial
integrals: 
\begin{eqnarray}
\fl
B_{ll^{\prime}}
=
{\ee^{\i\bigl(\delta_{l}(p)-\delta_1(p)\bigr)}\over p }
\int_R^{\infty}\!\!\int_R^{\infty}  \d r_1 \d r_2\,
{X_{p1}(r_1) \over r_1}\,
{\chi_{pl}(r_2)\over r_2}
 \Biggl(
g_{\E_{\rm i} l^{\prime}}(r_2,r_1)
+ 
g_{\E+\om l^{\prime}}(r_2,r_1) 
\Biggr).
\label{LongDistance1.5a}
\end{eqnarray}
For any realistic target the exact evaluation of these integrals
can be done by numerical means only, implying the computation of the 
regular $\chi_{pl}(r)$  and the irregular $X_{p1}(r)$ 
functions as well as of the Green's
functions $g_{\E_{\rm i} l^{\prime}}(r_2,r_1)$ and  
$g_{\E+\om l^{\prime}}(r_2,r_1) $.
To carry out the approximate integration one can substitute
these quantities with the leading terms of their
 asymptotic representations.
For $\chi_{pl}(r)$ and $X_{p1}(r)$ this results in 
$\chi_{pl}(r) \sim \sin\bigl(pr - \pi l/ 2 + \delta_l(p)\bigr)$ and
$X_{p1}(r) \sim -\ee^{\i\left(pr + \delta_1(p)\right)}$.
The leading terms in the expansions of the radial Green's functions are:
\begin{eqnarray}
\cases{
g_{\E+\om l^{\prime}}(r_2,r_1)
\sim
- {\i\over P}\,{\ee^{\i Pr_{>} }\over r_{>}}\,
{S_{l^{\prime}}(P)\ee^{\i Pr_{<}} - \ee^{-\i Pr_{<}}  \over r_{<}}\,,
\\
g_{\E_{\rm i}l^{\prime}}(r_2,r_1)
\sim
- {1\over p_i}\,
{\ee^{- p_i r_{>}} \over r_{>}}\,
{S_{l^{\prime}}(\i p_i )\ee^{-p_i r_{<}} - \ee^{p_i r_{<}}  \over r_{<} }\,.
}
\label{FunctionB2.2}
\end{eqnarray}
Here $P=\sqrt{2(\E+\om)}$, $p_i=\sqrt{-2\E_{\rm i}}$, 
$S_{l^{\prime}}(P)$  and $S_{l^{\prime}}(\i p_i)$ are the elements
of the scattering matrix. 
For the real argument $P$ this quantity reads as
$S_{l^{\prime}}(P)=\ee^{2\i \delta_{l^{\prime}}(P)}$.

By using the asymptotic formulae one carries out analytical evaluation 
of the integrals from (\ref{LongDistance1.5a}).
The final result (the leading term) for $\Im\,B_{ll^{\prime}}$ is as
follows:
\begin{eqnarray}
\fl
\Im\,B_{ll^{\prime}}
=&
{(-1)^{(l-1)/2} \over 2 X^4}
\Biggl[
\xi^{-1}\,{\xi^2 + 1 \over (\xi^2 - 1)^2} 
\nonumber\\
\fl
&
+
\xi^{-1}\,{\rm Re} \!
\left(
{S_{l^{\prime}}(P)S_l(p) \ee^{2\i (\xi+ 1) X}\over (\xi+ 1)^2} 
+
{S_{l^{\prime}}(P)\ee^{2\i \xi X} \over \xi^2 - 1} 
-{S_l(p)\ee^{2\i X}\over \xi+1}
\right) 
\nonumber\\
\fl
&
+
\xi_i^{-1}\,{\rm Im} { S_l(p) (\i\xi_i-1) \ee^{2\i X}\over (\xi_i^2+1)}
\Biggr]_
{l=1,3 \atop l^{\prime}=0,2},
\label{FunctionB3.3}
\end{eqnarray}
 where $X=pR$, $\xi = P/p$ and $\xi_i = p_i/p$.
It follows from the approximations formulated above, that 
formula (\ref{FunctionB3.3}) is valid if the photoelectron momentum $p$ 
and the ground state energy $\E_{\rm i}$  satisfy the conditions:
\begin{eqnarray}
pR > 1, \qquad
\sqrt{-2\E_{\rm i}}R > 1.
\label{FunctionB3.3a}
\end{eqnarray}

Letting $\delta_l(p) = \delta_{l^{\prime}}(P) \equiv 0$ 
in  (\ref{FunctionB3.3})
one finds the plane-wave Born limit of 
$\Im\,B_{ll^{\prime}}$. 
 
The presence of oscillatory terms 
on the right-hand side of (\ref{FunctionB3.3}) 
is physically clear. 
Indeed, in the limit $pR>1$ the wavelength of the photoelectron
is less than the size of the target, which is supposed to have a 
well-pronounced edge (e.g., the edge of an ionic core in a 
cluster). 
Thus, the electron experiences a diffraction at the edge when
leaving the core. 
This diffraction leads
to the oscillatory character of the cross section.
This phenomenon was discussed previously in connection with various 
processes involving metallic clusters and  fullerenes: 
the  radiative electron capture by metallic clusters 
\cite{ConneradeSolovyov_1996},
the electron scattering 
\cite{GerchikovSolovyovConneradeGreiner_1997},
the polarizational bremsstrahlung 
\cite{GerchikovSolovyov_1997}.

Carrying out obvious summations in (\ref{LongDistance1.5}), one writes
correction factor as follows:
\begin{eqnarray}
K(\om,\theta) 
&=
 E_0^2\, 
 \Bigl(\kappa_0(\om) +  \kappa_2(\om)\, P_2(\cos\theta) \Bigl),
\label{CorrectionFactor2.1}
\end{eqnarray}
where
\begin{eqnarray}
\fl
\kappa_0(\om)
=
-{1\over 3}
|\alpha(\om)|^2\,
\Im 
\left(
B_{10} + {4  \over 5}\, B_{12} + {4 \over 5}\, B_{32}
\right),
\qquad
\kappa_2(\om)
=
{2\over 3}
|\alpha(\om)|^2\,
\Im\, B_{32}.
\label{CorrectionFactor2.3}
\end{eqnarray}

Formula (\ref{CorrectionFactor2.1}) stresses the proportionality
of $K(\om,\theta)$ to the laser field intensity (which is proportional to
$E_0^2$) and  exhibits explicitly its dependence on the emission angle
$\theta$, 
which enters, as well as in (\ref{CS.4a}),
via the second order Legendre polynomial.

As a function of $\theta$ the modulus of $K(\om,\theta)$
attains its maximum either at $\theta=0^{\circ}$ 
(and $\theta=180^{\circ}$) or at $\theta=90^{\circ}$, 
depending on the sign of the product $\kappa_0(\om)\kappa_2(\om)$ and on 
the relative magnitudes of $\kappa_0(\om)$ and $\kappa_2(\om)$.
For further reference let us define the quantity
\begin{eqnarray}
\fl
|\kappa(\om)|_{\max}
&\equiv
\Bigr|\kappa_0(\om) +  \kappa_2(\om)\, P_2(\cos\theta) \Bigl|_{\max}
\nonumber\\
\fl
&=
\cases{ 
|\kappa_0(\om) +  \kappa_2(\om)| 
& 
if $\kappa_0(\om)\kappa_2(\om)>0$,
\\
|\kappa_0(\om)| +  |\kappa_2(\om)|/2
& 
if $\kappa_0(\om)\kappa_2(\om)<0$ and
$4|\kappa_0(\om)|>|\kappa_2(\om)|$,  
\\
|\kappa_2(\om)| -  |\kappa_0(\om)|
& 
if $\kappa_0(\om)\kappa_2(\om)<0$
and
$4|\kappa_0(\om)|<|\kappa_2(\om)|$.
}
\label{CorrectionFactor2.1a}
\end{eqnarray}
which allows one to estimate the laser field strength needed
to achieve a noticeable correction to the angular distribution
(see the discussion in section \ref{NumericalResults}).
   
Substituting (\ref{CorrectionFactor2.1}) into (\ref{CS.3}),
taking into account (\ref{CS.4a}) and expressing $P_2^2(\cos\theta)$
via  $P_2(\cos\theta)$ and $P_4(\cos\theta)$, one 
derives the following formula for the differential cross section 
within the lowest order of perturbation theory in
$\calU(\bfr,t)$:
\begin{eqnarray}
{\d \sigma \over \d \Omega} 
=
{\sigma_0 \over 4\pi}\,
\Bigl( 
a_0(\om) + a_2(\om)P_2(\cos\theta) + a_4(\om)P_4(\cos\theta)
\Bigr).
\label{Modification2.3}
\end{eqnarray}
The coefficients $a_{0,2,4}$ are related to 
the asymmetry parameter $\beta$ and to $\kappa_{0,2}$ via
\begin{eqnarray}
\fl
a_0 = 1+ E_0^2\left(\kappa_0 + {\beta\,\kappa_2 \over 5}\right),
\quad
a_2  = \beta 
+ 
 E_0^2\left[
\kappa_2 
+ 
\left(\kappa_0 + {2\kappa_2\over 7}\right)\beta\right],
\quad
a_4 ={18 \over 35}\,E_0^2\,\beta\,\kappa_2. 
\label{Modification2.9}
\end{eqnarray}

The right-hand side of (\ref{Modification2.3}) suggests that 
not only the profile of the angular distribution can be changed, 
but the magnitude
of the cross section, integrated over the emission angles, is scaled
by the factor $a_0$ which can be greater or lower than one depending
on the signs and magnitudes of $\beta$ and $\kappa_{0,2}$.

If the action of $\calU(\bfr,t)$ is ignored then
$a_0 \to 1$, $a_0 \to \beta$ and $a_4 \to 0$, and equation 
(\ref{Modification2.3}) reduces to (\ref{CS.4a}).

\section{Numerical results \label{NumericalResults}} 

Numerical analysis of the influence of the induced dipole moment
on the profile of the angular distribution was carried out for the
process of photodetachment of sodium cluster anions Na$_{N}^{-}$.
The results presented below in this section refer to  $N=8, 20, 40, 58$ and
$92$. 
The corresponding neutral clusters are spherically-symmetric
which makes applicable the theory described in section  \ref{Framework}.

The parameters of sodium clusters and their anions, 
used in our numerical calculations,
 are summarized in table \ref{NaN.Table}.
Let us comment on these data.

The ionic core radius was calculated using the standard relation
$R=r_sN^{1/3}$ with the Wigner-Seitz radius  $r_s$ set to 4 a.u. 
The values of electron affinities $\Ia$ are taken from   
a recent work by Kostko \cite{Kostko_disser} which contains, in particular, 
an excellent collection of reference data on various metal clusters.
The other data presented in the table refer to the 
polarizability of the core and the parameters of the plasmon peak.
We considered the dynamic dipole polarizability $\alpha(\om)$ of the clusters
within the framework of the resonance plasmon approximation 
(see, e.g., \cite{GerchikovSolovyov_1997,LesHouches}):
\begin{eqnarray}
\alpha(\om) 
= \alpha(0)\, { \om_{\rm s}^2 \over \om_{\rm s}^2 - \om^2 - \i\om\Gamma}.
\label{Numerical.1}
\end{eqnarray}
Here  $\alpha(0)$ is the static polarizability, $\om_{\rm s}$ is the 
surface plasmon energy, and 
$\Gamma$ is the linewidth of the plasmon resonance.

Due to the spill-out effect the static polarizability of a metallic
cluster of a radius $R$ exceeds the classical value $\alpha_{\rm cl}=R^3$, 
which characterizes a  metallic ball of the same radius. 
In the table we present the ratio 
$\alpha(0)/\alpha_{\rm cl}$ calculated within the RPA \cite{GuetJohnson1992}
and the  LDA \cite{Kurkina2001}) as well as the experimentally measured
values \cite{KnightClemengerHeerSaunders1985}. 
The $\om_{\rm s}$ values
were obtained with the help of the relation proposed in  
\cite{ReinersEllertSchmidtHaberland1995} to account for the 
size-dependence of the plasmon energy:
$\om_{\rm s} = \om_{\rm Mie}\bigl(1-1.5\delta/R\bigr)$, 
where $\om_{\rm Mie}=3.27$ eV is the Mie plasmon energy in a sodium
metallic sphere,
and $\delta=0.54$ \AA\, takes into account the spill-out effect.

\begin{table}
\caption{
Data on several spherically-symmetric sodium clusters:
the ionic core radius $R$;
the electron affinity $\Ia$ \cite{Kostko_disser},
$\alpha_{\rm cl} = R^3$ is the classical static polarizability,
$\alpha(0)$ stands for calculated \cite{GuetJohnson1992,Kurkina2001} 
and measured \cite{KnightClemengerHeerSaunders1985} 
static polarizability;
$\om_{\rm s}$ is the surface plasmon energy calculated  
according to \cite{ReinersEllertSchmidtHaberland1995},
$\Gamma$ is the width of the plasmon resonance peak 
\cite{MolinaWeinmannJalabert_2002}.}
\begin{indented} 
\item[]\begin{tabular}{@{}r|rrrrrrrr}
\br
          &  \centre{1}{$R$}   &  \centre{1}{$\Ia$}&  \centre{1}{$\alpha_{\rm cl}$}
          &  \centre{3}{$\alpha(0)/\alpha_{\rm cl}$} 
          &  \centre{1}{$\om_{\rm s}$} &  \centre{1}{$\Gamma$} \\
          & (a.u.)&   (eV)     & (a.u.)           & \cite{GuetJohnson1992}
          & \cite{Kurkina2001}& \cite{KnightClemengerHeerSaunders1985}
          &    (eV)       &   (eV)   \\
\br
Na$_8$    & 8.00  & 0.915 &  512 & 1.48 & 1.44 & 1.72 & 2.64 & 0.2 \\
Na$_{20}$ & 10.86 & 1.337 & 1280 & 1.41 & 1.37 & 1.67 & 2.81 & 0.4 \\
Na$_{40}$ & 13.68 & 1.509 & 2560 & 1.10 & 1.32 & 1.60 & 2.90 & 0.7 \\
Na$_{58}$ & 15.48 & 1.719 & 3712 & 1.24 & 1.22 &   -  & 2.95 & 0.4 \\
Na$_{92}$ & 18.06 & 1.847 & 5888 & 1.22 & 1.20 &   -  & 2.99 & 0.2 \\
\br
\end{tabular}
\label{NaN.Table}
\end{indented}
\end{table}

The size dependence of the width $\Gamma$ of the plasmon resonance
for neutral sodium clusters with $N$ up to several hundreds
was analyzed in 
\cite{YannouleasVigezziBroglia_1993} within the RPA scheme and in
\cite{MolinaWeinmannJalabert_2002} within a semiclassical approach.
These two methods, being consistent for $N>58$, deviate noticeably for 
lower $N$ values. 
The RPA calculations reproduce the $1/R$ scaling law which, is inadequate
when compared to the experimental data for $N\lesssim 58$ (see the discussion
in  \cite{YannouleasVigezziBroglia_1993}). 
The semiclassical calculations 
 \cite{MolinaWeinmannJalabert_2002} seem to be consistent with the
experimental data in the whole region $N=10\dots 100$.
The values of $\Gamma$, presented in the table, were deduced from 
figure 1 in  \cite{MolinaWeinmannJalabert_2002}.

It is seen from the table, that for all $N$ the plasmon energy exceeds 
the electron affinity.
This fact justifies the choice of cluster anions in a view of the 
present study.
Indeed, the inequality $\om_{\rm s}>\Ia$ leads to the additional 
enhancement
of $K(\om,\theta)$ owing to the pronounced increase of the modulus of the 
dynamic polarizability in the region $\om \approx \om_{\rm s}$.

To conclude the discussion on the parameters of the chosen targets 
let us note, that both criteria from (\ref{FunctionB3.3a}) 
can be easily met for Na$_{N}^{-}$.
The first inequality fails only in the narrow near-threshold
region $\om- \Ia\ll \Ia$ but is well fulfilled for 
$\om \approx \om_{\rm s}$.
The validity of the second condition one proves 
by introducing $\E_{\rm i} = -\Ia$.

\begin{figure}[ht]
\centering
\includegraphics[clip,scale=0.45]{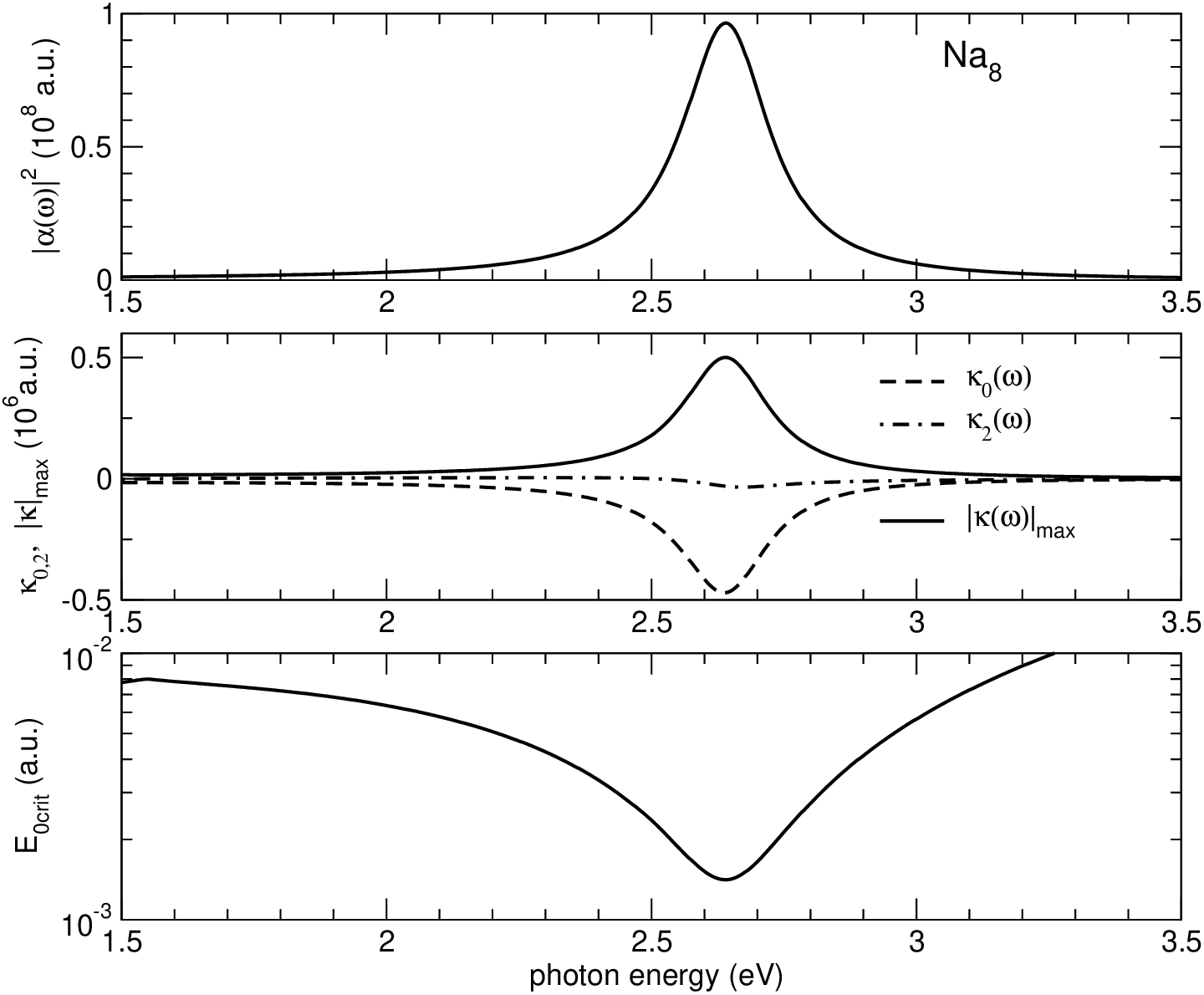}
\\
\vspace*{0.4cm}
\includegraphics[clip,scale=0.45]{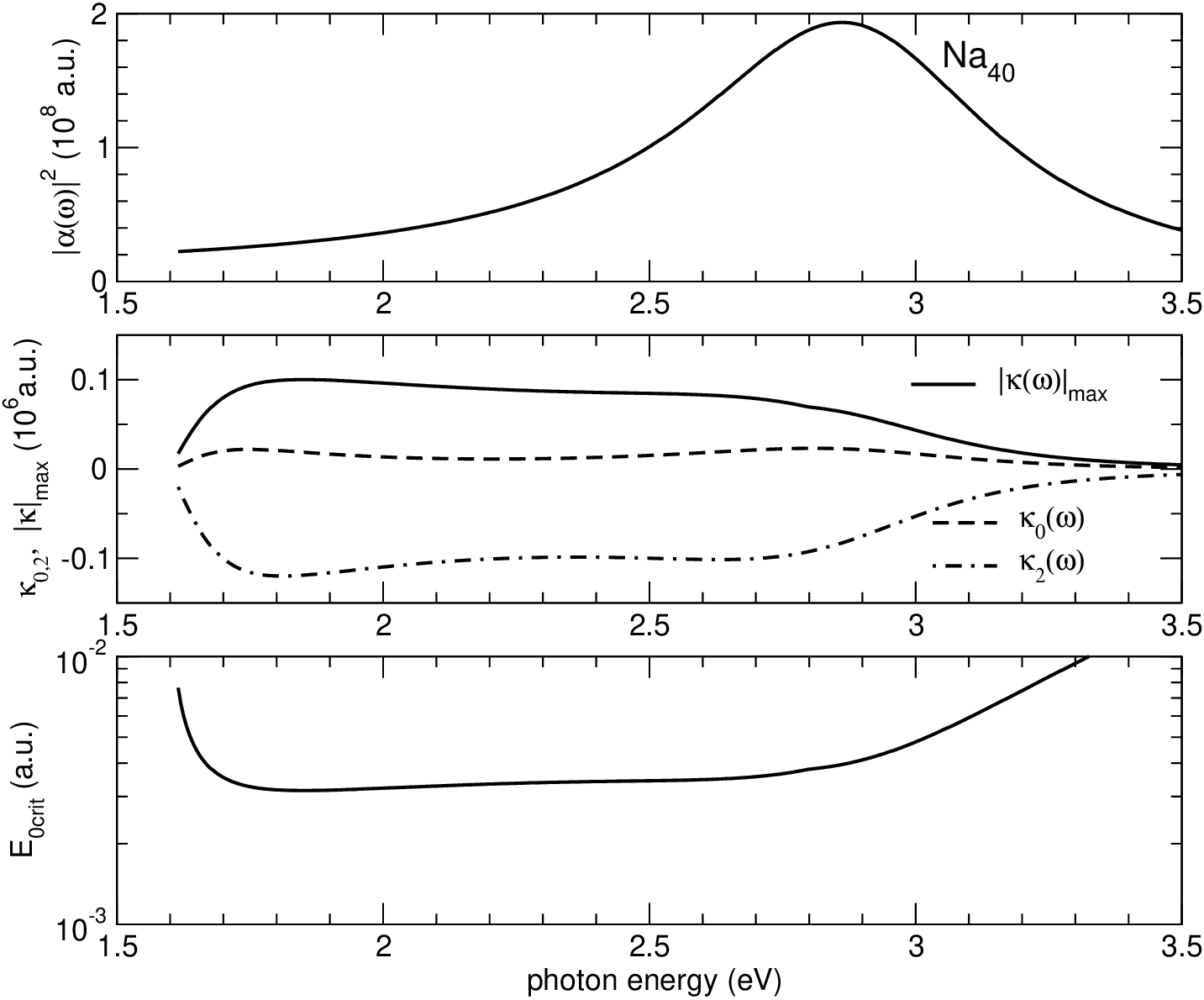}
\caption{
Calculation of the correction term for Na$_{8}^{-}$ and for Na$_{40}^{-}$ as
indicated.
\\ 
In each graph the upper panel represents the 
dependence of $|\alpha(\om)|^2$  on $\om$;
the middle panel \--  
dependences $\kappa_{0,2}(\om)$ and $|\kappa(\om)|_{\max}$,
eqs. (\ref{CorrectionFactor2.3})-(\ref{CorrectionFactor2.1a});
the lower panel \-- dependence of 
$E_{0\rm crit}(\om)$, eq. (\ref{CorrectionFactor2.1b}).
Parameters $\om_{\rm s}$ and $\Gamma$ are 
as indicated in table \ref{NaN.Table}.
The data refer to the $\alpha(0)/R^3$ ratio equal to 1.45 for Na$_{8}$ 
and to 1.30 for Na$_{40}$ .
}
\label{na08_40.fig}
\end{figure}

\begin{figure}[ht]
\centering
\includegraphics[clip,scale=0.45]{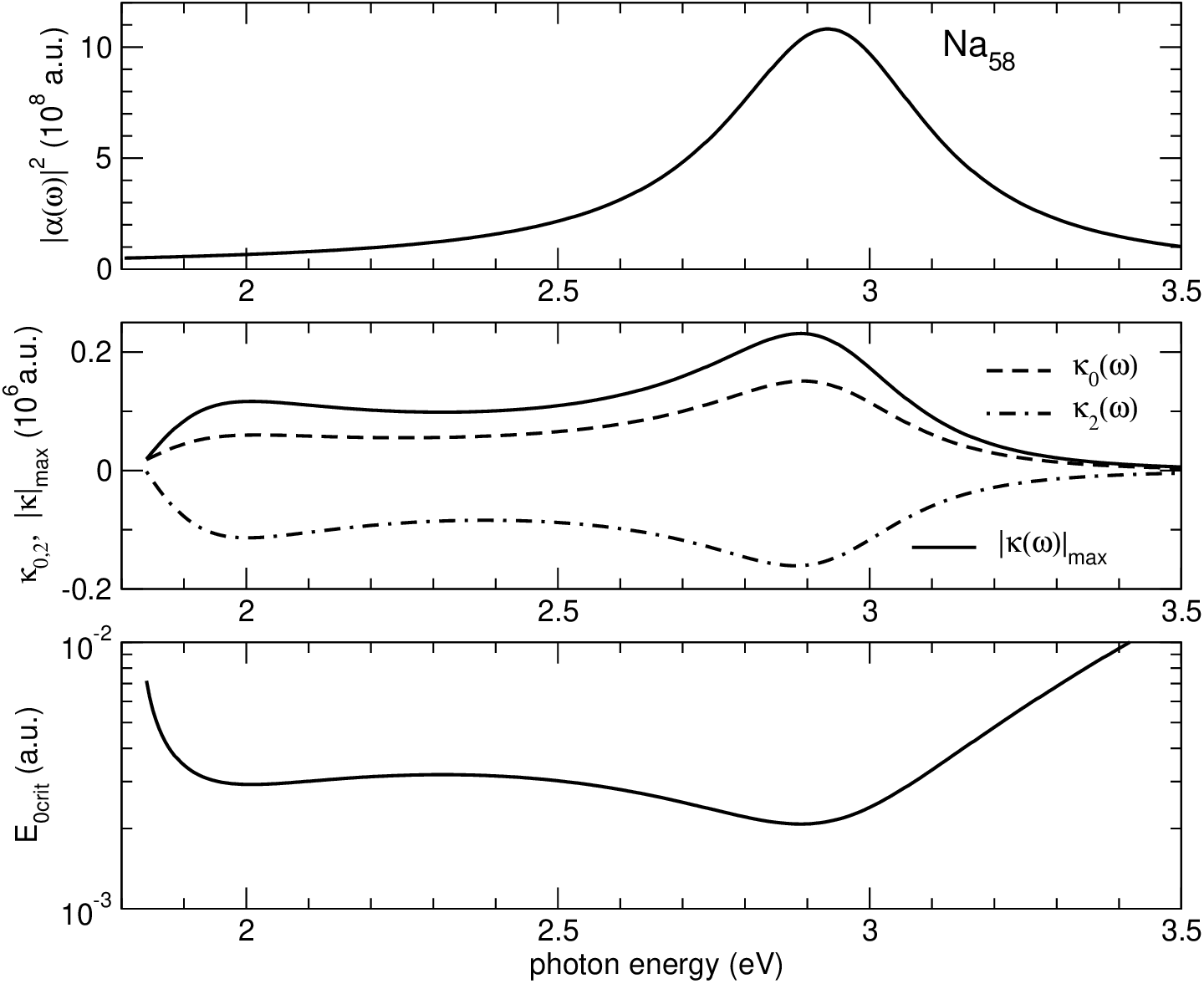}
\\
\vspace*{0.4cm}
\includegraphics[clip,scale=0.45]{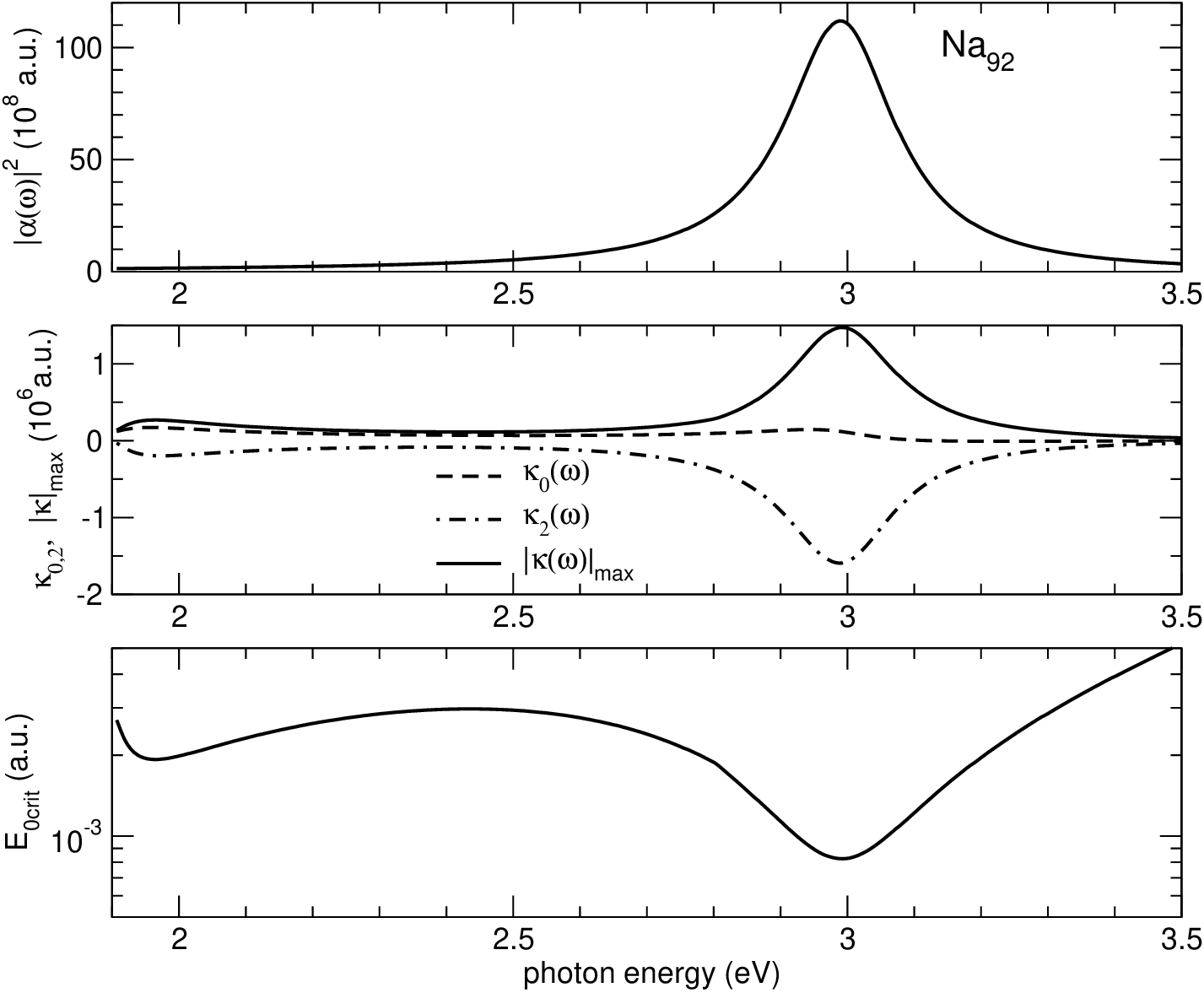}
\caption{
Same as in figure \ref{na08_40.fig} but for Na$_{58}^{-}$ and
Na$_{92}^{-}$.
For both clusters the ratio $\alpha(0)/R^3$ is set to 1.20.
}
\label{na58_92.fig}
\end{figure}

The results of calculation of the quantities relevant to the 
correction term $K(\om,\theta)$
are presented in figures \ref{na08_40.fig} and \ref{na58_92.fig}. 
For each anion the calculations were performed for the photon 
energies above the ionization threshold $\Ia$ and 
in the vicinity of the plasmon peak.
The latter manifests itself as a resonance in the dependence 
of $|\alpha(\om)|^2$ on $\om$, as it is seen from the upper panel 
in each graph.
The dynamic polarizabilities were calculated via (\ref{Numerical.1}) with
the parameters $\om_{\rm s}$ and $\Gamma$ as indicated in table 
\ref{NaN.Table}.
Given the differences between the calculated and experimentally measured
values of the static polarizabilities $\alpha(0)$,
as well as the absence of the experimental data for Na$_{58}$ and  Na$_{92}$,
for the ratio $\alpha(0)/R^3$ we used the values  close to the theoretical
data (see the figures captions). 

The dependences of the coefficients $\kappa_0(\om)$ and $\kappa_2(\om)$ 
on $\om$ are presented in the middle panels.
They were computed from (\ref{CorrectionFactor2.3}) and 
(\ref{FunctionB3.3}).
To calculate the right-hand side of the latter formula beyond the 
plane-wave Born approximation
one must know the scattering phaseshifts $\delta_{l}(p)$  and 
$\delta_{l}(P)$.
To determine the phaseshifts we solved numerically the phase equation
(see, e.g., \cite{Babikov,Calogero}) for an electron moving in 
a local potential.
The model potential, which was used to describe the interaction between 
the photoelectron and the neutral residue, accounted for the following 
three terms:
(a) the short-range potential due to the Coulomb interaction 
with the core's electrons and the ionic jellium core,
(b) the exchange-correlation term, which was treated
within the Gunnarsson-Lundqvist approximation 
\cite{GunnarssonLundqvist_1976}, 
(c) the polarization potential was considered in the 
form  $U_{\rm pol}(r)=-\alpha(0)/2(r^2+R^2)^2$ for $r>R$ and 
$U_{\rm pol}(r)=0$ if otherwise.
Inclusion of the polarization term is important for an adequate 
description the low-energy electron scattering from metallic clusters   
(e.g., \cite{BernathEtAl_1995}).  
A more rigorous treatment accounts for  the many-body correlations, 
which may be considered, for example, within the Dyson equation scheme 
\cite{DescourtFarineGuet_2000}.   
However, for the purpose of our paper it is sufficient to treat
the polarization potential in the simple form described above.

Comparing the middle panels in the graphs from figures 
\ref{na08_40.fig} and \ref{na58_92.fig}
one notices that the behaviour of $\kappa_{0,2}(\om)$
(and of $|\kappa(\om)|_{\max}$ as well) exhibits 
some general trends along with the specific features determined by the
dynamic and geometrical properties of an individual  target.
Common to all clusters is the existence of extrema for $\kappa_{0,2}(\om)$
at $\om \approx \om_{\rm s}$, which is due to the factor 
$|\alpha(\om)|^2$, see (\ref{CorrectionFactor2.3}).
The extremum can be clearly defined (see the curves
$\kappa_0(\om)$ for Na$_{8}$ and $\kappa_2(\om)$ for Na$_{92}$ 
which have distinct maxima)
or less pronounced (both curves for  Na$_{58}$)
or barely seen (the case of Na$_{40}$).  
The variation in the shape of the extremum, as well as the behaviour 
of the curves in the
regions below and above the plasmon resonance, can be understood
by analyzing the $\om$-dependence
of the right-hand sides in 
(\ref{CorrectionFactor2.3}) and  (\ref{FunctionB3.3}).
The non-oscillatory part of the latter depends on the photon 
energy mainly
through the factor $(pR)^{-4} = 4R^{-4}(\om-\Ia)^{-2}$.
This leads to 
$\kappa_{0,2}(\om)\propto |\alpha(\om)|^2/(\om-\Ia)^{2}$.
Therefore, a symmetric resonance profile of  
$|\alpha(\om)|^2$ is distorted due to the factor $(\om-\Ia)^{-2}$.
For a powerful and narrow resonance (as in the case of Na$_{92}$) 
the distortion is comparatively small, whereas for a wide resonance 
the disproportion between the low- and the high-$\om$ shoulders may
completely change the behaviour of the curve in the vicinity of the resonance,
as it happens for Na$_{40}$.    
In either case the factor $(\om-\Ia)^{-2}$ results in a 
faster decrease of $|\kappa_{0,2}(\om)|$ with $\om$ beyond the resonance.
In the region $\Ia < \om < \om_{\rm s}$ the curves 
 $\kappa_{0,2}(\om)$, being enhanced  by the factor,
exhibit a non-monotonous oscillatory character
(most clearly seen in Na$_{40}$, Na$_{58}$ and Na$_{92}$ graphs).
This is due to the second and the third terms on the right-hand side 
of (\ref{FunctionB3.3}), 
which contain the oscillating factors $\exp(\i pR)$ and  $\exp(\i PR)$,
dependent on the cluster size,  
and the factors $S_{l}(p)=\exp\bigl(\i\delta_l(p)\bigr)$,
$S_{l^{\prime}}(P)=\exp\left(\i\delta_l(P)\right)$,
dependent on the phaseshifts.

The solid curves in the middle panels of each graph represent the 
dependences  $|\kappa(\om)|_{\max}$ defined in (\ref{CorrectionFactor2.1a}).
For a given photon energy the quantity $E_0^2\, |\kappa(\om)|_{\max}$ stands 
for the maximum value of 
$|K(\om,\theta)|$ within the interval $\theta = [0^{\circ},180^{\circ}]$.
It was already mentioned, that the correction term must satisfy 
the condition $|K(\om,\theta)| < 1$.
Thus, taking into account that 
$|K(\om,\theta)| \leq E_0^2 |\kappa(\om)|$,
one can introduce a critical value $E_{0\, {\rm crit}}(\om)$, 
which defines the upper boundary for the laser field strength consistent 
with the perturbative approach adopted in this paper:
\begin{eqnarray}
E_{0\, {\rm crit}}(\om) = \left(|\kappa(\om)|_{\max}\right)^{-1/2}.
\label{CorrectionFactor2.1b}
\end{eqnarray}
The corresponding intensity of the laser field (in W/cm$^2$) can be calculated
from the relation 
$I_{0\, {\rm crit}}\approx 3.3\times 10^{16} E_{0\, {\rm crit}}^2$, where
$E_{0\, {\rm crit}}$ is measured the atomic units. 

In the limit $E_0 \ll E_{0\, {\rm crit}}(\om)$ a strong inequality
$|K(\om,\theta)| \ll 1$ is valid, so that the photoionization process
is not affected by the induced dipole moment.
The influence of the latter can become pronounced for 
$E_0 \lesssim E_{0\, {\rm crit}}(\om)$ resulting in 
$|K(\om,\theta)| \sim 1$.

\begin{table}
\caption{
The minima $\tom$ of the functions $E_{0\, {\rm crit}}(\om)$
(bottom panels in the graphs in figures \ref{na08_40.fig} and \ref{na58_92.fig}),
the values of $\kappa_{0,2}$ and 
$|\kappa|_{\max}$ in these points,
the critical strength of the laser field  
$E_{0\, {\rm crit}}(\tom)=\left(|\kappa(\tom)|_{\max}\right)^{-1/2}$ and the
corresponding intensity $I_{0\rm crit}(\tom)$,
the ponderomotive energy  $E_{0\rm crit}^2(\tom)/4\tom^2$;
$E_{\rm a}$ is the internal anionic field.
}
\footnotesize\rm
\begin{tabular}{@{}l|crrccccc}
\br
         & $\tom$ & \centre{1}{$\kappa_0(\tom)$} & \centre{1}{$\kappa_2(\tom)$}
         & $|\kappa(\tom)|_{\max}$ & $E_{0\rm crit}(\tom)$
         & $I_{0\rm crit}(\tom)$ & $E_{\rm a}$ & $E_{0\rm crit}^2(\tom)/4\tom^2$ \\
         &    (eV) &   ($10^{6}$)  &   ($10^{6}$) &    ($10^{6}$) &   ($10^{-3}$ a.u.)
         &($10^{11}$W/cm$^2$)      &($10^{-3}$ a.u.) &  ($10^{-3}$ eV) \\
\br
Na$_8$   &  2.64 &  -0.470 &  -0.031 &  0.501 &  1.41  &  0.66  &  17.5  &  1.44 \\
Na$_{20}$&  2.73 &   0.021 &   0.022 &  0.043 &  4.82  &  7.67  &  30.8  &  15.7 \\ 
Na$_{40}$&  2.72 &   0.022 &  -0.100 &  0.078 &  3.58  &  4.23  &  37.0  &  8.64 \\
Na$_{58}$&  2.89 &   0.151 &  -0.160 &  0.231 &  2.08  &  1.43  &  44.9  &  2.61 \\
Na$_{92}$&  2.98 &   0.128 &  -1.589 &  1.461 &  0.83  &  0.23  &  50.1  &  0.39 \\    
\br
\end{tabular}
\label{NaN.Table2}
\end{table}

Dependences $E_{0\, {\rm crit}}(\om)$ are plotted in the 
lower panels in figures  \ref{na08_40.fig} and \ref{na58_92.fig}. 
For all clusters this function attains its 
minimum in the point $\tom\approx \om_{\rm s}$ in the vicinity of 
the plasmon resonance. 
The values of $\tom$, as well as of other quantities calculated in this point, 
are presented in table \ref{NaN.Table2}.
It is instructive to compare the critical field $E_{0\, {\rm crit}}(\tom)$
with the strength $E_{\rm a}$ of the internal anionic field.
The latter can be related to the ionization
potential $\Ia$ through $E_{\rm a} = (2\Ia)^{3/2}$.
It is seen that $E_{\rm a}$ exceeds $E_{0\, {\rm crit}}(\tom)$ by
the order of magnitude.
The strong inequality $E_{\rm a} \gg E_{0\, {\rm crit}}(\tom)$,
being in accordance with the first condition from (\ref{conditions}),
justifies the approximation adopted in this paper to neglect the 
modification of the ground 
state orbitals due to the laser field.
The last column in the table contains the values of 
ponderomotive energy  of the photoelectron 
in the critical laser field.
Comparing the ponderomotive shifts and the kinetic energies 
$\E = \tom - \Ia \sim 1$ eV,  one notices that the latter is larger 
by several orders of magnitude.
Therefore, the second condition in (\ref{conditions}) is also well 
fulfilled.
 
Hence, one can expect, that for the intensities 
$I_0 \lesssim I_{0\, {\rm crit}}(\tom) \sim 10^{11}$ W/cm$^2$ 
the modification of the spectral-angular distribution
due to the action of the field of the induced dipole moment can be quite
pronounced.
To carry out the corresponding quantitative estimates one can use 
the formulae 
obtained within the framework of perturbation theory.
%
\begin{figure}[ht]
\centering
\includegraphics[clip,scale=0.43]{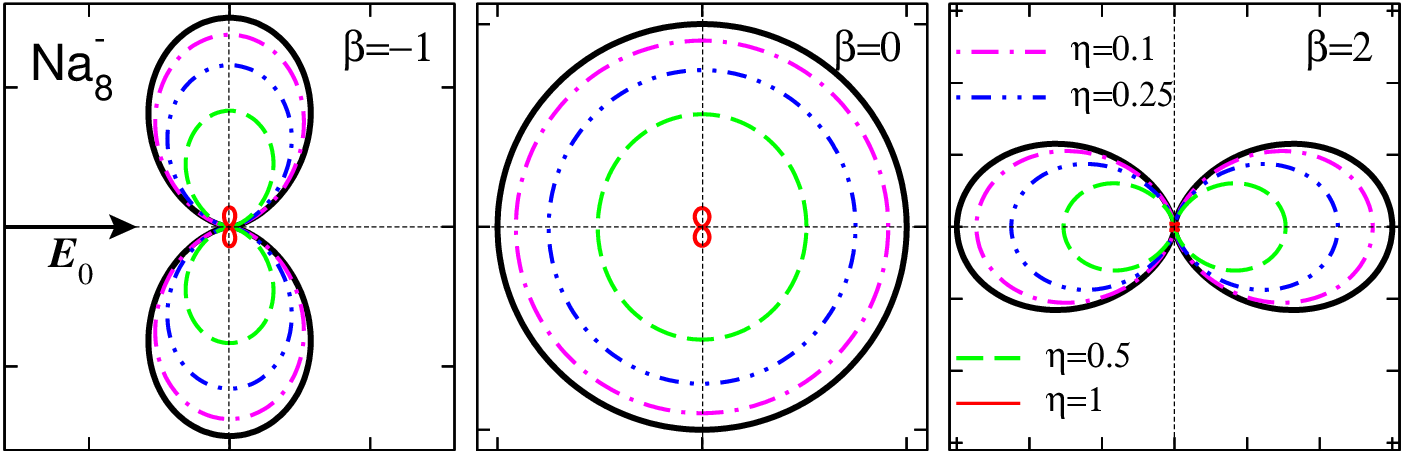}
\\
\includegraphics[clip,scale=0.43]{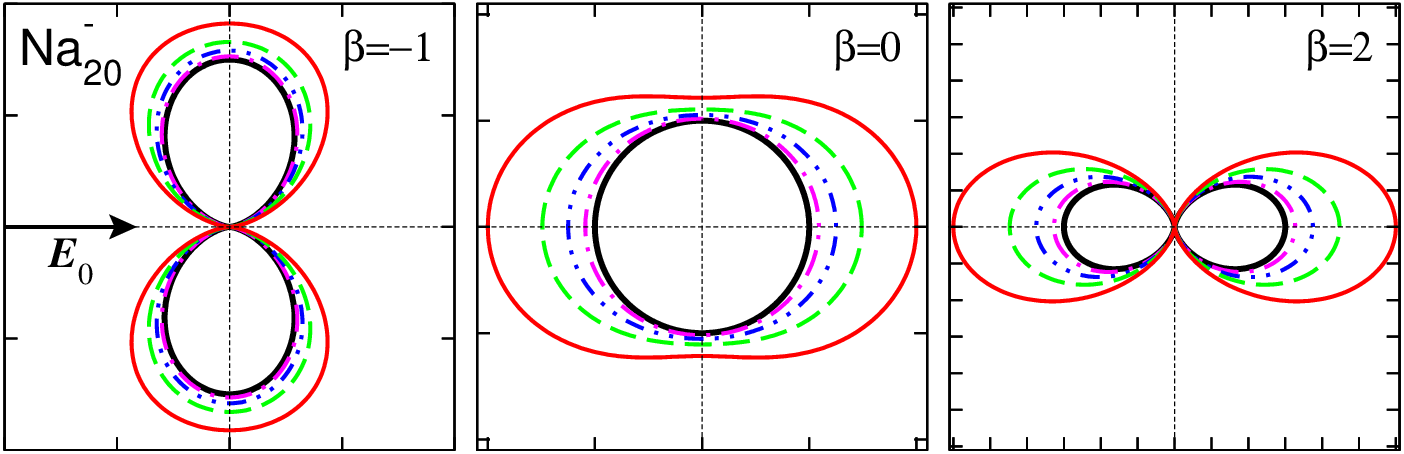}
\\
\includegraphics[clip,scale=0.43]{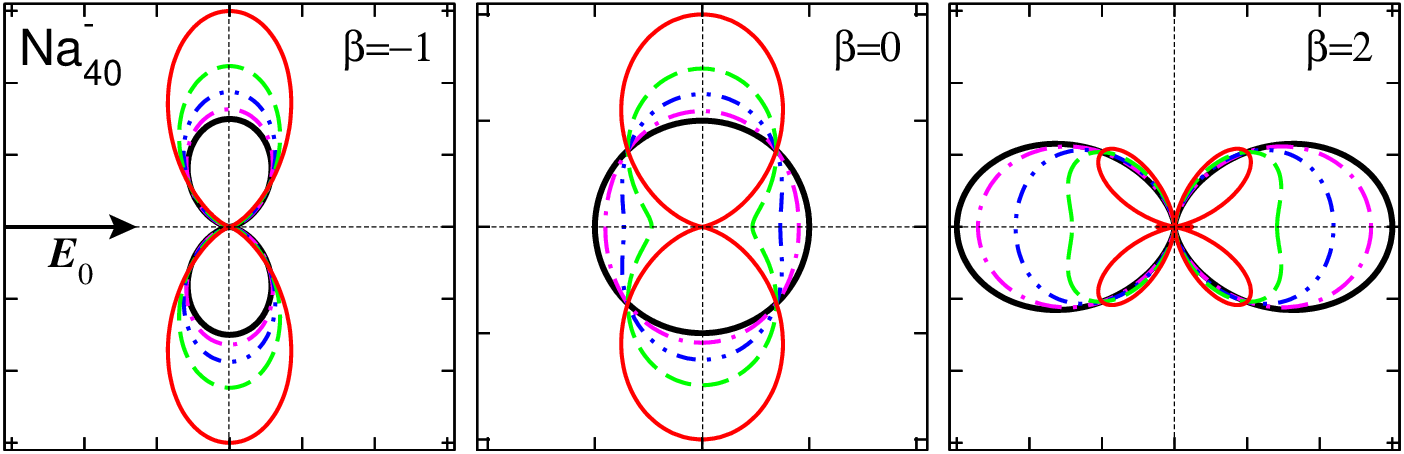}
\\
\includegraphics[clip,scale=0.43]{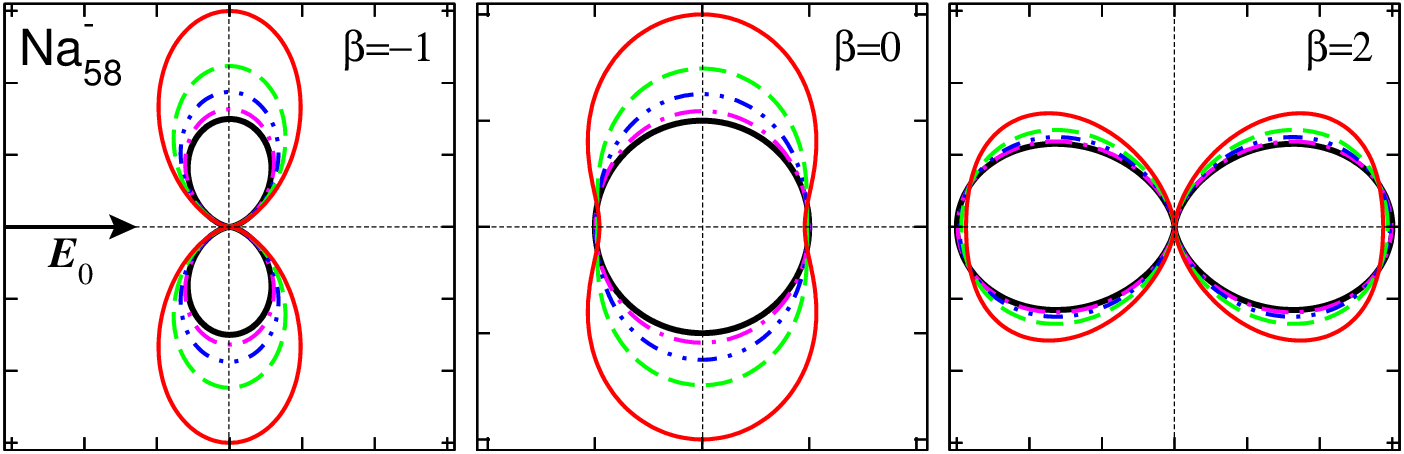}
\\
\includegraphics[clip,scale=0.43]{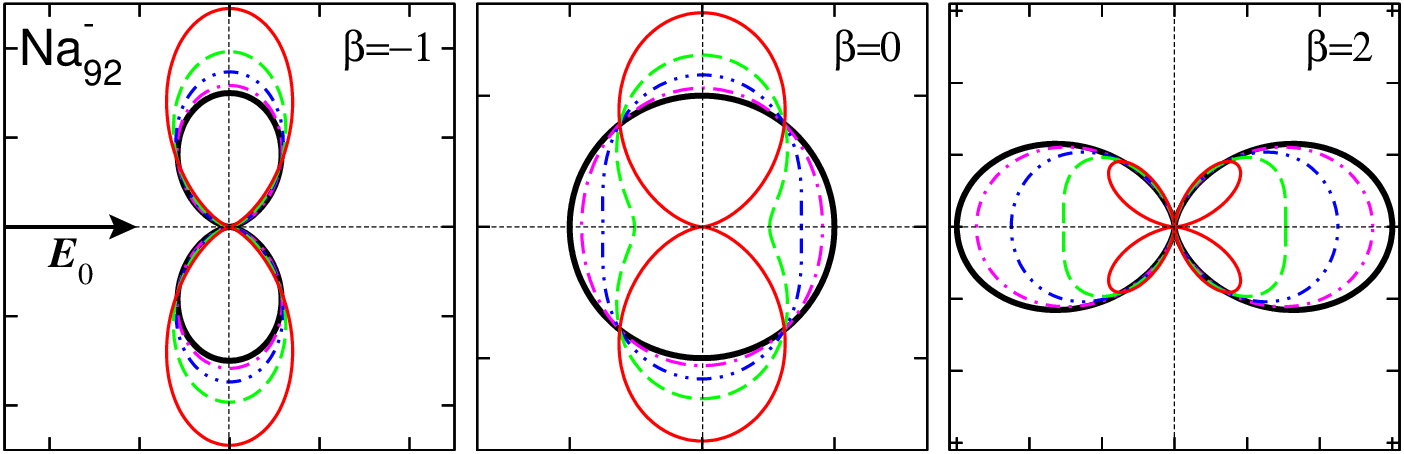}
\caption{
Profiles $\calF_{\beta}(\theta)$ of the angular distribution 
(\ref{profile.1}) calculated for 
different sodium cluster anions and for the three values of asymmetry 
parameter $\beta$  (as indicated), and for several values of 
the ratio $\eta = I_0/ I_{0\, {\rm crit}}$
(see the legend in the graph $\beta=2$ for Na$_8^{-}$). 
The data refer to the photon energies $\tom$ indicated in table
\ref{NaN.Table2}.
Thick black curves in each graph represent the field free distribution
$F_{\beta}(\theta)$, eq. (\ref{profile.0}).
The length of line segment connecting the origin and a given point on 
the curves is equal to the value of $\calF_{\beta}(\theta)$  or
$F_{\beta}(\theta)$  (in absolute units) in the corresponding direction.
The emission angle $\theta$ is measured with respect to $\bfE_0$, which
is shown in the graphs with $\beta=-1$.
In each graph the spacing between ticks is equal to 1 for both directions.
}
\label{profiles.fig}
\end{figure}

The calculated profiles of the angular distributions are presented in
figure  \ref{profiles.fig}.
In the present paper we have not made an attempt to compute the 
angular asymmetry 
parameter $\beta(\tom)$ for sodium anion clusters 
(neither have we found its values in the literature).
Instead, for each target we carried out the calculations for the 
three values, $\beta=-1, 0, 2$, aiming to demonstrate that the 
modification of the profile takes place for {\it any}  $\beta$ from 
the allowable interval $[-1,2]$. 

In each graph from figure \ref{profiles.fig} the thick solid curve 
represents the unperturbed profile 
\begin{eqnarray}
F_{\beta}(\theta) \equiv 1+\beta P_2(\cos\theta)
=
\cases{
{3\over 2}\sin^2\theta  & for $\beta=-1$, \\
1                       & for $\beta=0$, \\
3\cos^2\theta           & for $\beta=2$. 
}
\label{profile.0}
\end{eqnarray}
The emission angle $\theta$ is measured with respect to 
the vector $\bfE_0$, which is indicated in the graphs 
with $\beta=-1$.
Other curves correspond to the modified profiles
\begin{eqnarray}
\fl
\calF_{\beta}(\theta) 
\equiv
F_{\beta}(\theta)
\left(1+E_0^2\biggl(\kappa_0 + \kappa_2P_2(\cos\theta)\Bigr)\right)
&=
F_{\beta}(\theta)
\left(1
+\eta\,
{\kappa_0 + \kappa_2P_2(\cos\theta)\over |\kappa|_{\max}} 
\right)
\label{profile.1}
\end{eqnarray}
(see (\ref{CS.3}), (\ref{CorrectionFactor2.1}) and 
(\ref{CorrectionFactor2.1b}))
calculated for several values of the ratio 
$\eta = I_0/ I_{0\, {\rm crit}} \leq 1$.
For each anion the data refer to the photon energy $\tom$.

Figure \ref{profiles.fig} demonstrates, that even for comparatively 
low intensities  (corresponding to $\eta \gtrsim 10^{-1}$) 
the changes in the profiles are already visible. 
For $\eta \leq 1$ they become much more pronounced.
It is seen that variation of the profile depends, 
both qualitatively and quantitatively, 
on the value of $\beta$  and on the choice of the target anion which enters 
via the coefficients $\kappa_0$ and $\kappa_2$.

In the case $\beta=-1$ the field-free profile $F_{\beta}(\theta)$ is
proportional to $\sin^2\theta$ yielding 
$F_{-1}(0^{\circ}) =0$ as the minimum 
and  $F_{-1}(90^{\circ}) =1.5$ as the maximum values 
within the interval $\theta=[0^{\circ},90^{\circ}]$.
The modified profile can be written as  
$\calF_{-1}(\theta) \propto \sin^2\theta\, (a+b\cos^2\theta )$
with $a=1+ \eta(2\kappa_0-\kappa_2)/2|\kappa|_{\max}$ and
$b=3\eta\kappa_2/2|\kappa|_{\max}$. 
One easily shows, that for $|a| < |b|$ the function 
$\calF_{-1}(\theta)$ acquires additional extremum. 
However, this inequality cannot be met for the considered targets 
(see the values of $\kappa_0$, $\kappa_2$ and $|\kappa|_{\max}$ in
 table \ref{NaN.Table2}) and for $\eta \leq 1$.
Therefore, the pattern of the modified angular distribution 
is similar to that obtained without the influence of the induced dipole.
Namely, the strongest emission occurs in the directions perpendicular to 
the field $\bfE_0$ and there is no emission for 
$\theta=0^{\circ}$ and $180^{\circ}$.
On the whole,  the distortion increases with $\eta$, attaining the largest
values at $\theta=90^{\circ}$. 
The trend of the largest distortion depends on the sign of $\kappa_0$: 
if $\kappa_0 > 0$ (as in the cases of all anions but Na$_8$) then 
$\calF_{-1} > F_{-1}$ whereas
negative $\kappa_0$ leads to a shrinking of the modified profiles
(as it happens for Na$_8^{-}$).

For $\beta=0$ the unperturbed angular distribution is isotropic.
The modified profile reads as
$\calF_{0}(\theta) \propto a+b\cos^2\theta$, with $a$ and $b$ defined
as above.
The way the profile evolves with the increase of $\eta$ to a great 
extent depends on the sign of $\kappa_2$.
For $\kappa_2>0$ the emission along $\pm \bfE_0$ 
dominates over the emission in the perpendicular directions (see the 
 case of Na$_{20}^{-}$), whereas for negative $\kappa_2$ the perpendicular
direction becomes preferable.
The graphs from the central column in figure \ref{profiles.fig} 
demonstrate,
that for $\eta \approx 1$ the distortion can become dramatic both in 
form and in magnitude.
 
In the limit $\beta=2$ the modified profile behaves as
$\calF_{2}(\theta) \propto \cos^2\theta\, (a+b\cos^2\theta )$.
The additional extrema appear if the following two conditions, 
$ab<0$ and $|a| < 2|b|$, are met.
These conditions can be satisfied in the limit $\eta\approx 1$ for 
all considered anions but Na$_{20}^{-}$.
Most explicitly it is seen in Na$_{40}^{-}$ and Na$_{92}^{-}$ graphs, 
where the  curves with $\eta=1$
represent the four-petal angular distributions with the maximum emission
at $\theta=45^{\circ}$ and $135^{\circ}$.
The same pattern characterizes Na$_{8}^{-}$, although in this case 
it is hardly visible due to
a strong decrease in the magnitude of $\calF_{2}(\theta)$ for $\eta=1$. 

\begin{figure}[ht]
\centering
\includegraphics[clip,scale=0.575]{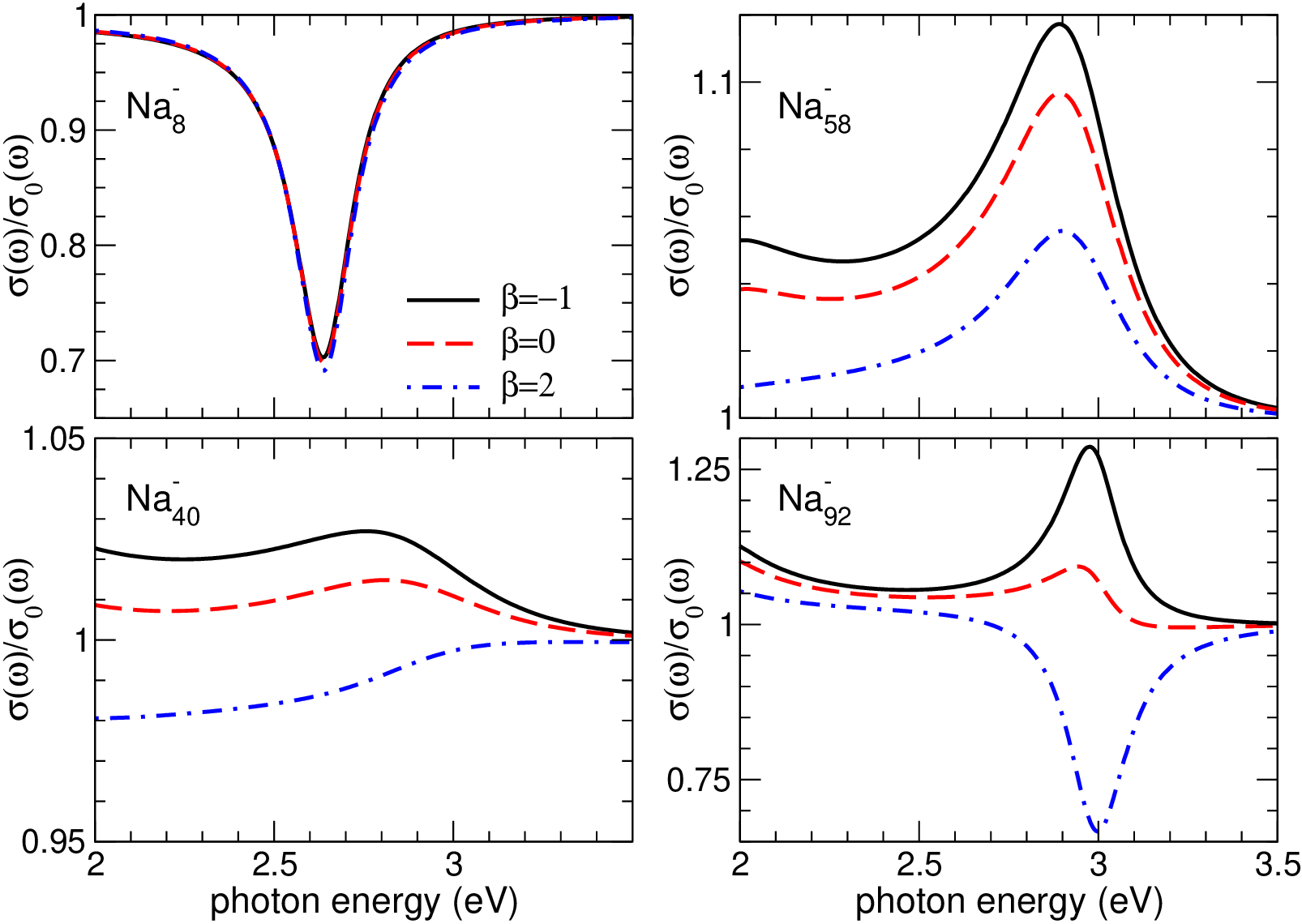}
\caption{
Ratio $\sigma(\om)/\sigma_0(\om)$ (see (\ref{spectra.1}))
calculated for different sodium cluster anions and for the three values 
of asymmetry parameter $\beta$  (as indicated in the top left graph).
The data refer to the laser field intensity 
$I_0 = 2.1\times10^{10}$ W/cm$^2$.
}
\label{spectra.fig}
\end{figure}

Finally, let us briefly discuss the influence of the induced dipole moment 
on the cross section integrated over the emission angles.
Recalling (\ref{Modification2.3}) and (\ref{Modification2.9}),
one finds the following formula for the ratio of the cross sections
with, $\sigma(\om)$, and without,  $\sigma_0(\om)$, 
account for the correction:
\begin{eqnarray}
{\sigma(\om) \over \sigma_0(\om)} 
=
1+ E_0^2\left(\kappa_0(\om) + {\beta(\om)\,\kappa_2(\om) \over 5}\right).
\label{spectra.1}
\end{eqnarray}
Figure \ref{spectra.fig} presents the ratios $\sigma(\om)/\sigma_0(\om)$
versus $\om$.
For each anion the ratio was calculated using the dependences 
$\kappa_0(\om)$ and $\kappa_0(\om)$, presented in
figures \ref{na08_40.fig} and \ref{na58_92.fig}, and 
for three values of the asymmetry 
parameter (as indicated).
The intensity of the laser field was set to 
$I_0 = 2.1\times10^{10}$  W/cm$^2$, which corresponds to the field 
strength $E_0=8\times10^{-4}$ a.u. 
Comparing the latter value to the critical field $E_{0\, {\rm crit}}(\om)$ 
(see figures \ref{na08_40.fig} and \ref{na58_92.fig})
one notices, that for all targets and in the whole range of $\om$ 
the parameter $\eta = E_0^2/E_{0\, {\rm crit}}^2(\om)$ is less than one.
Thus, formula (\ref{spectra.1}) is applicable for the quantitative estimate 
of the correction due to the induced moment.

As one can see from figure \ref{spectra.fig}, the largest 
deviation (up to $\approx 30\%$) of $\sigma(\om)$ from $\sigma_0(\om)$ 
occurs for  Na$_{8}^{-}$ and Na$_{92}^{-}$ in the vicinities of 
the plasmon resonances, where $\eta$ reaches its maximum values:
$\eta \approx 0.57$ for Na$_{8}^{-}$ and $\eta \approx 0.95$ for 
Na$_{92}^{-}$.
In the case of Na$_{92}^{-}$ a strong inequality $|\kappa_2|\gg |\kappa_0|$
is valid (see figure \ref{na58_92.fig} or/and table \ref{NaN.Table2}).
Therefore, the magnitude of the correction term from
the right-hand side of (\ref{spectra.1}) becomes sensitive to the choice
of the asymmetry parameter.
As a result, the ratio $\sigma(\om) /\sigma_0(\om)$ strongly varies with
$\beta$.
On the contrary, for Na$_{8}^{-}$ the relation 
$|\kappa_2|\ll |\kappa_0|$ is valid, leading to a weak dependence
of the ratio on $\beta$.

For the anions  Na$_{40}^{-}$ and Na$_{58}^{-}$ the largest values  of 
$\eta$ equal to 0.05 and 0.15, correspondingly. 
Hence, the intensity $I_0$ is small compared to 
$I_{0\, {\rm crit}}(\om) \propto E_{0\, {\rm crit}}^2(\om)$.
As a result, the deviation of $\sigma(\om)$ from the field-free 
value $\sigma_0(\om)$ is less pronounced.

\section{Summary}

In summary, we have demonstrated that the dipole moment, induced
at the residue by the incoming laser field, can strongly influence
the photoionization process. 
The additional time-dependent long-range field due to the induced moment, 
acting on the escaping electron, modifies the
spectral-angular and spectral distributions of photoelectrons.

The effect itself, as well as its quantitative treatment, is quite 
sensitive to the choice of the target and to the parameters of the laser
field.
We have demonstrated, that for a strongly polarizable target 
(e.g., metallic cluster anions) strong modifications 
of the characteristics of a single-photon ionization process in 
the vicinity of the plasmon 
resonance (typically,  $\om_{\rm s}\approx 2-3$ eV) 
can be achieved by applying laser fields of comparatively 
low intensities $I_0 \sim10^{10}-10^{11}$  W/cm$^2$.

From a theoretical viewpoint, the advantage of this regime is that 
the field of induced dipole moment can be treated
perturbatively, whereas the influence of the laser field on the photoelectron
(a ponderomotive effect) and on the ground state of the target 
can be ignored.
We have shown, that even in this weak-field regime one can expect 
dramatic changes in the profile of angular distribution of
photoelectrons as well as in the $\om$ dependence of the photoionization
cross section. 
To a great extent, these changes depend on the parameters of the target, 
like its radius, the static and dynamic polarizabilities, 
the energy of the plasmon resonance peak and the affinity (or, the 
ionization potential in the case of a neutral target). 
Additional variety can be obtained by changing the intensity and the 
frequency of the laser field.

The analysis, which we have carried out, is based on simple models
adopted for the description of the photodetachment process (the plasmon
resonance approximation) and of the interaction of the escaping 
electron with the static field of the residue (via the LDA + long-range
polarization potential).
We have also restricted ourselves to the case of spherically-symmetric targets.
A more accurate quantitative treatment must include more sophisticated
approaches, which take into account the many-body correlation effects
intrinsic to both the photodetachment and the electron escape.
However, these will leave untouched the physics behind the basic phenomenon 
discussed in our paper.

It is possible to study the predicted phenomena 
by means of modern experimental techniques. 
The latter include the production and manipulation of the beams of 
cluster anions of various types and sizes, the energy and angle resolved
photoelectron spectroscopy, the availability of the laser fields of 
various intensity, frequency and pulse duration 
\cite{KostkoEtAl_2007,Kostko_disser}. 
We believe, that with all these components at hand, soon it will become
possible to test the theory against experimental data.

\ack
We are grateful to Roman Polozkov for the help in numerical calculation
of the ground states of sodium clusters, and to Steven Manson and 
Bernd von Issendorff for the helpful discussions.

This work was supported by the  European Commission within the 
Network of Excellence project EXCELL (project number 515703).


\end{document}